\documentclass[conference]{IEEEtran}
\IEEEoverridecommandlockouts
\usepackage{cite}
\usepackage{amsmath,amssymb,amsfonts}
\usepackage{algorithmic}
\usepackage{graphicx}
\usepackage{tabularx}
\usepackage{textcomp}
\usepackage{xcolor}
\usepackage{stfloats}
\usepackage{multirow}
\usepackage{epstopdf}

\def\BibTeX{{\rm B\kern-.05em{\sc i\kern-.025em b}\kern-.08em
    T\kern-.1667em\lower.7ex\hbox{E}\kern-.125emX}}
\begin{document}

\title{A 3D Non-stationary MmWave Channel Model for Vacuum Tube Ultra-High-Speed Train Channels }

\author{Yingjie Xu\textsuperscript{1,2}, Kai Yu\textsuperscript{3}, Li Li\textsuperscript{4}, Xianfu Lei\textsuperscript{4,1}, Li Hao\textsuperscript{4}, Cheng-Xiang Wang\textsuperscript{1,2,*}\\
\textsuperscript{1}National Mobile Communications Research Laboratory, School of Information of Science and Engineering, 
\\Southeast University, Nanjing 210096, China.\\
\textsuperscript{2}Purple Mountain Laboratories, Nanjing 211111, China.\\
\textsuperscript{3}China Railway Eryuan Engineering Group Co. Ltd, Chengdu, Sichuan 610031, China.\\
\textsuperscript{4}School of Information Science and Technology, Southwest Jiao Tong University, Chengdu 610031, China.\\
\textsuperscript{*}Corresponding Author: Cheng-Xiang Wang\\
Email: yjxu@seu.edu.cn, ekyukai@qq.com,\\ 
\{ll5e08, xflei, lhao\}@home.swjtu.edu.cn, chxwang@seu.edu.cn
}

\maketitle

\begin{abstract}
As a potential development direction of future transportation, the vacuum tube ultra-high-speed train (UHST) wireless communication systems have newly different channel characteristics from existing high-speed train (HST) scenarios. In this paper, a three-dimensional non-stationary millimeter wave (mmWave) geometry-based stochastic model (GBSM) is proposed to investigate the channel characteristics of UHST channels in vacuum tube scenarios, taking into account the waveguide effect and the impact of tube wall roughness on channel. Then, based on the proposed model, some important time-variant channel statistical properties are studied and compared with those in existing HST and tunnel channels. The results obtained show that the multipath effect in vacuum tube scenarios will be more obvious than tunnel scenarios but less than existing HST scenarios, which will provide some insights for future research on vacuum tube UHST wireless communications.
\end{abstract}


\section{Introduction}
With the social development and population growth, transportation is developing rapidly. In the future, vacuum tube transportation systems can overcome the limitations of current wheel-rail transportation environment such as air resistance and train wheel-rail resistance on the speed of trains, and the train speed can reach thousands of kilometers per hour \cite{ref1}, becoming an important development direction of HST transportation systems. For vacuum tube UHST train-to-ground wireless communication systems, the applications of the fifth generation (5G) wireless communication networks, which only support up to 500 km/h mobility \cite{ref2}, are not enough. Accordingly, this will promote research on ultra high mobility in the sixth generation (6G) wireless communication networks \cite{ref3}.

There are many channel models which can well reflect the wireless channel propagation characteristics in existing HST and tunnel scenarios \cite{ref4,ref5,ref6,ref7}. In \cite{ref8} a deterministic channel model for HST scenarios was proposed, and channel small-scale fading characteristics were investigated. A mmWave massive MIMO GBSM for HST communication systems was shown in \cite{ref9}, where the author studied HST channel statistical properties. The channel statistical properties in tunnel scenarios were investigated in \cite{ref10} by proposing a three-dimensional (3D) non-stationary GBSM. However, these models cannot be directly used for vacuum tube UHST channels. The vacuum and narrow space environment have newly different effects on its wireless channel, and the ultra high speed will bring larger Doppler frequency and more fast handovers \cite{ref11}. In \cite{ref12}, a propagation graph channel model for vacuum tube UHST scenarios was proposed, and several channel properties were analyzed, such as multipath, K factor, and channel capacity. However, it focused on channels without considering mmWave technologies that can provide high data rate transmissions to communication systems. 
The  general 3D non-stationary 5G wireless channel model in \cite{ref13} can reflect channel characteristics of most scenarios. However, due to the assumption of random cluster distribution and missing consideration of the waveguide effect in channels, it cannot be directly applied to the vacuum tube UHST channel.

To the best of author’s knowledge, non-stationary mmWave vacuum tube UHST channel models are still missing in the literature. To fill the research gaps, a 3D non-stationary mmWave GBSM for vacuum tube UHST scenarios is proposed in this paper. In the model, it is assumed that the scattering clusters are distributed on the inner wall of the tube, and then the vacuum tube UHST channel properties are studied. 

The remainder of this paper is organized as follows. A 3D non-stationary mmWave GBSM is proposed in Section II and channel statistical properties are derived in Section III. Section IV illustrates simulation results and discussions about UHST channels. Finally, conclusions are shown in Section V.  
\section{A 3D Non-Stationary mmWave MIMO GBSM for Vacuum Tube UHST Scenarios}
\subsection{Description of Vacuum Tube UHST Communication Network Architecture}
\begin{figure}[tb]
	\centerline{\includegraphics[width=0.48\textwidth]{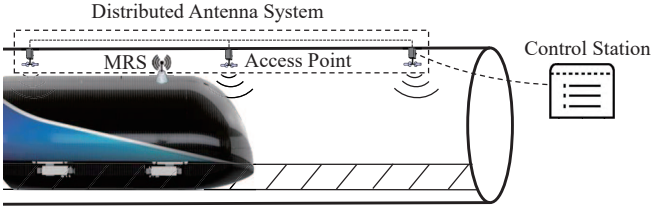}}
	\caption{A UHST network architecture for vacuum tube scenarios.}
	\label{UHST_Newwork_Architecture}
\end{figure}

The communication network architecture for vacuum tube UHST scenarios is shown in Fig.~\ref{UHST_Newwork_Architecture}. In communication systems, it is considered to adopt technologies, such as distributed antenna system (DAS), radio over fiber (RoF), and mobile relay station (MRS) \cite{ref10}. The access points (APs) are used and fixed on the top of the metal tube inner wall to form the DAS. They are connected by the RoF, and finally connected to the control station (CS) at the station. A small MRS is installed on surface of the train to reduce high penetration loss of the signal entering train compartment.
By distributing the DAS and the MRS in communication systems, the propagation space between base stations and the train is divided into multiple parts. This paper will aim to investigate the channel between AP and MRS.
\subsection{The mmWave MIMO GBSM for Vacuum Tube UHST Scenarios}
\begin{figure}[tb]
	\centerline{\includegraphics[width=0.5\textwidth]{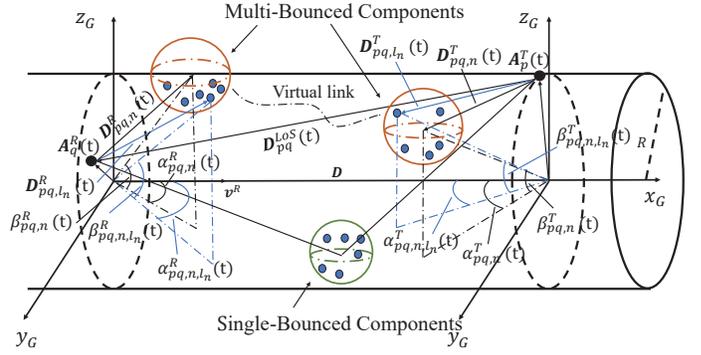}}
	\caption{A 3D non-stationary mmWave MIMO GBSM for vacuum tube UHST scenarios.}
	\label{channel_model}
\end{figure}

Let us consider a MIMO system with $Q$ and $P$ antenna elements at the receiver (Rx) and transmitter (Tx) side and let ${\bf{A}}_p^T(t) $ and ${\bf{A}}_q^R(t) $ denote 3D position vectors of the $p$th transmit antenna and the $q$th receive antenna. 
Based on the 5G general channel model \cite{ref13}, a 3D non-stationray GBSM is proposed, where the channel propagation environment is characterized as a 3D cylindrical model with radius $R$. The Cartesian coordinate system is used to describe the location of the Tx and the Rx. It is assumed that the Rx (MRS) on train moves towards the Tx (AP), as shown in Fig.~\ref{channel_model}. 

\subsubsection{Channel Impulse Response}
The complete channel matrix is given by ${\bf{H}} = {[PL \cdot SH \cdot BL \cdot OL]^{\frac{1}{2}}} \cdot {{\bf{H}}_s}$, where $PL$ represents the path loss, $SH$ represents the shadowing, $BL$ represents the blockage loss, and $OL$ represents the oxygen and molecular absorption loss. Widely used path loss model and shadowing model are shown in \cite{ref14}. The blockage loss is caused by train and obstacles in vacuum tube UHST scenarios and its model is taken from \cite{ref15} here. The oxygen and molecular absorption loss model for mmWave can be found in \cite{ref16}. Note that the parameter values in above models should be measured additionally and will be different from those measured in the standard atmosphere because of the vacuum environment in UHST channels. 

The small-scale fading can be denoted as a complex matrix ${{\bf{H}}_s} = {[{h_{pq}}(t,\tau )]_{P \times Q}}$, where ${h_{pq}}(t,\tau )$ is the channel impulse response (CIR) between ${\bf{A}}_p^T(t) $ and ${\bf{A}}_q^R(t) $ that consists of line-of-sight (LoS) component and non-line-of-sight (NLoS) components. The NLoS components include single-bounced (SB) and multi-bounced (MB) components. The propagation environment between Tx and Rx is abstracted by effective clusters that characterize the first and last bounce of the channel. 
Assuming there are a total of $N(t)$ effective clusters on tube wall at time $t$ in channel, and $L_n$ is the time-variant number of rays within ${\rm{Cluste}}{{\rm{r}}_n}$. Note that each ray in each cluster should have its own power and delay to support higher spectral resolution in models, which is the difference between mmWave channels and conventional channels. 
The CIR ${h_{pq}}(t,\tau )$ can be expressed as\cite{ref13}
\begin{equation}
\begin{split}
{h_{pq}}(t,\tau )&= \underbrace {h_{pq}^{LoS}(t) \cdot \delta (\tau  - {\tau ^{LoS}}(t))}_{LoS} \\
&+ \underbrace {\sum\limits_{n = 1}^N {\sum\limits_{{l_n} = 1}^{{L_n}} {h_{pq,n,{l_n}}^{NLoS}(t) \cdot \delta (\tau  - {\tau _n}(t) - {\tau _{{l_n}}}(t))} } }_{NLoS}.
\end{split}
\label{CIR}
\end{equation}
In (\ref{CIR}), ${\tau ^{LoS}}(t) $ is the delay of LoS component, ${\tau _n}(t) $ is the delay of ${\rm{Cluste}}{{\rm{r}}_n}$ and the delay ${\tau _{{l_n}}}(t) $ of the $l_n$th rays within ${\rm{Cluste}}{{\rm{r}}_n}$ is taken into account to support mmWave scenarios. Moreover, to model high mobility and non-stationarity of UHST channels, all parameters in the model are time-variant. Suppose the train speed vector at the Rx is ${{\bf{v}}^R}$, and initial distance vector between the Rx and Tx is ${\bf{D}} $. The key channel parameters are defined in Table I.

For the LoS component, the complex channel $h_{pq}^{LoS}(t) $ can be written as
\begin{equation}
h_{pq}^{LoS}(t) = \sqrt {\frac{{{K_{pq}}(t)}}{{{K_{pq}}(t) + 1}}} {e^{ - \frac{{j2\pi D_{pq}^{LoS}}}{\lambda }}}{e^{j2\pi f_{pq}^{LoS}(t) \cdot t}}
\end{equation}
where $f_{pq}^{LoS}(t) $ can be calculated as 
\begin{equation}
\begin{aligned}
f_{pq}^{LoS}(t)& = \frac{1}{\lambda }\frac{{\left\langle {{\bf{D}}_{pq}^{LoS}(t),{{\bf{v}}^R}} \right\rangle }}{{\left\| {{\bf{D}}_{pq}^{LoS}(t)} \right\|}}.
\end{aligned}
\end{equation} 
The delay ${\tau ^{LoS}}(t) $ can be expressed as
\begin{equation}
{\tau ^{LoS}}(t) = \frac{{\left\| {{\bf{D}}_{pq}^{LoS}(t)} \right\|}}{c} 
\end{equation} 
where $\left\|  \cdot  \right\|$ calculates the Frobenius norm and ${\bf{D}}_{pq,n,{l_n}}^{NLoS}(t) $ can be calculated as 
\begin{equation}
{\bf{D}}_{pq}^{LoS}(t) = {\bf{A}}_q^R(t) - {\bf{A}}_p^T(t). 
\end{equation}

For NLoS components, the complex channel $h_{pq,n,{l_n}}^{NLoS}(t)$ can be written as 
\begin{equation}
\begin{split}
h_{pq,n,{l_n}}^{NLoS}(t)& = \sqrt {\frac{{{P_{n,{l_n}}}(t)}}{{{K_{pq}}(t) + 1}}} {e^{j({\varphi _{n,{l_n}}} - \frac{{2\pi D_{pq,n,{l_n}}^{NLoS}(t)}}{\lambda })}}\\
&\times {e^{j2\pi f_{pq,n,{l_n}}^{NLoS}(t) \cdot t}} 
\end{split}
\end{equation}
where ${\varphi _{n,{l_n}}} $ is initial phase,  ${D}_{pq,n,{l_n}}^{NLoS}(t) $ and $f_{pq,n,{l_n}}^{NLoS}(t) $ can be calculated as
\begin{equation}
D_{pq,n,{l_n}}^{NLoS}(t) = \left\| {{\bf{D}}_{pq,n,{l_n}}^T(t)} \right\| + \left\| {{\bf{D}}_{pq,n,{l_n}}^R(t)} \right\| + {\widetilde \tau _n}(t) \cdot c
\end{equation}
\begin{equation}
\begin{aligned}
f_{pq,n,{l_n}}^{NLoS}(t) = \frac{1}{\lambda }\frac{{\left\langle {{\bf{D}}_{pq,n,{l_n}}^R(t),{{\bf{v}}^R}} \right\rangle }}{{\left\| {{\bf{D}}_{pq,n,{l_n}}^R(t)} \right\|}}
\end{aligned}
\end{equation} 
where ${\widetilde \tau _n}(t) \cdot c $ represents the virtual link distance and ${\widetilde \tau _n}(t)  $ is a virtual delay. The distance vectors ${\bf{D}}_{pq,n,{l_n}}^R(t) $ and ${\bf{D}}_{pq,n,{l_n}}^T(t) $ are determined by the AoAs and AoDs of $l_n $th rays, as shown in (9) and (10).
\begin{equation}
\begin{split}
{\bf{D}}_{pq,n,{l_n}}^R(t) = \frac{R}{{\sqrt {1 - {{\cos }^2}\beta _{pq,n,{l_n}}^R(t){{\cos }^2}\alpha _{pq,n,{l_n}}^R(t)} }}\\
{\left[ \begin{array}{l}
	\cos \beta _{pq,n,{l_n}}^R(t)\cos \alpha _{pq,n,{l_n}}^R(t)\\
	\cos \beta _{pq,n,{l_n}}^R(t)\sin \alpha _{pq,n,{l_n}}^R(t)\\
	\sin \alpha _{pq,n,{l_n}}^R(t)
	\end{array} \right]^T} + {\bf{D}} - {\bf{A}}_q^R(t) 
\end{split}
\end{equation}
\begin{equation}
\begin{split}
{\bf{D}}_{pq,n,{l_n}}^T(t) = \frac{R}{{\sqrt {1 - {{\cos }^2}\beta _{pq,n,{l_n}}^T(t){{\cos }^2}\alpha _{pq,n,{l_n}}^T(t)} }}\\
{\left[ \begin{array}{l}
	\cos \beta _{pq,n,{l_n}}^T(t)\cos \alpha _{pq,n,{l_n}}^T(t)\\
	\cos \beta _{pq,n,{l_n}}^T(t)\sin \alpha _{pq,n,{l_n}}^T(t)\\
	\sin \alpha _{pq,n,{l_n}}^T(t)
	\end{array} \right]^T} - {\bf{A}}_q^T(t).
\end{split}
\end{equation} 
 Similarly, the distance vectors ${\bf{D}}_{pq,n}^R(t)$ and ${\bf{D}}_{pq,n}^T(t)$ of ${\rm{Cluste}}{{\rm{r}}_n}$ can be calculated by replacing $ \alpha _{pq,n,{l_n}}^R(t)$, $\beta _{pq,n,{l_n}}^R(t)$ in (9) with $\alpha _{pq,n}^R(t)$, $  \beta _{pq,n}^R(t)$ and replacing $ \alpha _{pq,n,{l_n}}^T(t)$, $\beta _{pq,n,{l_n}}^T(t)$ in (10) with $\alpha _{pq,n}^T(t)$, $  \beta _{pq,n}^T(t)$.
 \begin{table*}[tb]
 	\centering
 	\caption{Definition of Key Parameters.}
 		\begin{tabular}{|l|l|}			

			\hline
 			\textbf{Parameters}&\textbf{Definition}\\
 			\hline
 			${\bf{D}}_{pq}^{LoS}(t)$ & 3D distance vector between ${\bf{A}}_p^T(t)$ and ${\bf{A}}_q^R(t) $of the LoS component\\
 			\hline
 			${\bf{D}}_{pq}^{NLoS}(t)$ & 3D distance vector between ${\bf{A}}_p^T(t)$ and ${\bf{A}}_q^R(t)$ of the NLoS components
 			\\\hline
 			${\bf{D}}_{qp,n}^R(t) $, ${\bf{D}}_{qp,n}^T(t) $ & 3D distance vectors between ${\rm{Cluste}}{{\rm{r}}_n}$ and the Rx (Tx) array center
 				\\\hline
 			${\bf{D}}_{pq,n,{l_n}}^R(t) $,${\bf{D}}_{pq,n,{l_n}}^T(t) $ & 3D distance vectors between the $l_n$th rays within ${\rm{Cluste}}{{\rm{r}}_n}$ and the Rx (Tx) array center
 				\\\hline
 			${\bf{D}} $ & initial distance vector between Tx and Rx
 				\\\hline
 			$\alpha _{pq,n}^R(t)$, $\beta _{pq,n}^R(t) $ & azimuth and elevation angles between  ${\rm{Cluste}}{{\rm{r}}_n}$ and the Rx array center 
 				\\\hline
 			$\alpha _{pq,n}^T(t) $, $\beta _{pq,n}^T(t) $ & azimuth and elevation angles between  ${\rm{Cluste}}{{\rm{r}}_n}$ and the Tx array center 
 				\\\hline
 			$\alpha _{pq,n,{l_n}}^R(t) $, $\beta _{pq,n,{l_n}}^R(t) $ & 
        	azimuth and elevation angles between the $l_n$th rays within ${\rm{Cluste}}{{\rm{r}}_n}$ and the Rx array center
        		\\\hline
        	$\alpha _{pq,n,{l_n}}^T(t) $,$\beta _{pq,n,{l_n}}^T(t) $ & azimuth and elevation angles between the $l_n$th rays within ${\rm{Cluste}}{{\rm{r}}_n}$ and the Tx array center
        	 	\\\hline
        	
 			${\bf{A}}_q^R(t) $, ${\bf{A}}_p^T(t) $ & 3D position vectors of the $q$th antenna at Rx and the $p$th antenna at Tx
 			\\\hline
 		    $\textbf{v}^R$&3D velocity vector of receive array
 			\\\hline
 			${K_{pq}}(t) $ & Rician factor
			\\\hline
 	        $f_{pq}^{LoS}(t) $, $f_{pq,n,{l_n}}^{NLoS}(t) $ & Doppler frequency between ${\bf{A}}_p^T(t) $ and ${\bf{A}}_q^R(t) $ of the LoS (NLoS) component
 	        		\\\hline
 	        ${{N}(t)} $ & the total number of effective clusters at time $ t$ 
 	        			\\\hline
 	        ${P_{n,{l_n}}}(t)$ & the normalized power of $l_n$th rays within ${\rm{Cluste}}{{\rm{r}}_n}$	
 	        \\\hline		
 	        $ \lambda $ & the wavelength of the signal 
 	        	\\\hline

 		\end{tabular}

 		\label{tab1}
 \end{table*}
\subsubsection{Cluster Evolution in Time Domains for UHST channel}
The cluster evolution in time domains for UHST channel is achieved by updating cluster information and geometric characteristics of channels. Here we use the birth and death process \cite{ref9} to describe the time evolution of clusters, given appropriate cluster survival and recombination rates. 

For survived clusters, firstly, which clusters are survived should be determined. Let ${P_T}(\Delta t) $ denote the survival probability of a cluster after $\Delta t$ and its calculation are given in \cite{ref13}. Then, the position vector of receiving antenna is updated as
\begin{equation}
{\bf{A}}_q^R(t + \Delta t) = {\bf{A}}_q^R(t) + {{\bf{v}}^R}\Delta t.
\end{equation}
Similarly, the distance vectors in (5), (9)$\sim$(12) need to be updated accordingly. Next, the delay of ${\rm{Cluste}}{{\rm{r}}_n}$ is updated as
\begin{equation}
\begin{split}
&{\tau _n}(t + \Delta t) \\
&= \frac{{\left\| {{\bf{D}}_{pq,n,{l_n}}^R(t + \Delta t)} \right\| + \left\| {{\bf{D}}_{pq,n,{l_n}}^T(t + \Delta t)} \right\|}}{c}+ {\widetilde \tau _m}(t + \Delta t) 
\end{split}
\end{equation}
where the virtual delay ${\widetilde \tau _m}(t + \Delta t) $ at time $t + \Delta t $ is calculated as ${\widetilde \tau _n}(t + \Delta t) = {e^{ - \frac{{\Delta t}}{\varsigma }}}{\widetilde \tau _n}(t) + (1 - {e^{ - \frac{{\Delta t}}{\varsigma }}})X $. The random variable $X $ and ${\widetilde \tau _n} $ have the same distribution but are independent of each other and $\varsigma$ is a scenario-dependent parameter\cite{ref13}. Finally, the mean power of $l_n$th rays within ${\rm{Cluste}}{{\rm{r}}_n}$	need to be updated as [10]
\begin{equation}
{\widetilde P_{n,{l_n}}}(t + \Delta t) = {\widetilde P_{n,{l_n}}}(t)\frac{{3{\tau _n}(t) - 2{\tau _n}(t + \Delta t) + {\tau _{{l_n}}}}}{{{\tau _n}(t) + {\tau _{{l_n}}}}}. 
\label{update_power}
\end{equation}
It should be noted that the updated power in (\ref{update_power}) should be normalized before being substituted into (6).

For new cluster generations, the number of new clusters generated at a stationary interval should be determined at first. In the fully enclosed tube, the channel keyhole effect \cite{ref17} related to waveguide will cause the multipath components in signal to change with distance. As the distance between the Rx and Tx increases, the multipath components will experience more similar channel fadings. Moreover, it has been confirmed that the scattering surface roughness ${\sigma _h}$ will affect the scattering and reflection loss of the signal, and thus the number of multipaths reaching the Rx. By considering above phenomenons in vacuum tube channel, it is assumed that the number of new clusters generated at time $t$ follows a Poisson distribution, and its mean value is
\begin{equation}
E[{N_{new}}(t)] = \frac{{{\lambda _G}}}{{{\lambda _R}}}(1 - {P_T}(t))(1 - \frac{{\left\| {{{\bf{D}}^{LoS}}(t)} \right\|}}{D})  \frac{{{\rho _s}}}{{{\rho _{s0}}}} 
\end{equation}
where ${\rho _{s0}}$ is the scattering coefficient when the roughness ${\sigma _h} = 0$. The scattering coefficient ${\rho _{s}}$ is calculated as \cite{ref18}
\begin{equation}
{\rho _s} = {e^{( - 8{{(\frac{{\pi {\sigma _h}\cos (\overline \beta  )}}{\lambda })}^2})}} 
\end{equation}
where ${\bar \beta }$ is the mean elevation angle of the incident ray. After the number of new cluster determined, key parameters for new cluster need to be given. Here, the calculation of the ray power $P_{n,{l_n}}$ is referenced in \cite{ref18}.
The number of rays in cluster follows the Poisson distribution and the angle parameters follow the Von Mises distribution \cite{ref10}. The virtual delay ${\tau _n}$ follows the exponential distribution \cite{ref19}.

\section{Statistical Properties}
In this section, several typical statistical properties of the mmWave channel model for vacuum tube UHST scenarios will be derived.
\subsection{The Time-Variant Transfer Function}
The time-variant transfer function ${ H_{pq}}(t,f) $ is the Fourier transform of the CIR ${h_{pq}}(t,\tau )$ relative to $\tau  $, which can be expressed as 
\begin{equation}
\begin{split}
{H_{pq}}(t,f) &= \int\limits_{ - \infty }^\infty  {h_{pq}}(t,\tau ){e^{ - j2\pi \tau f}}d\tau  \\
&= h_{pq}^{LoS}(t){e^{ - j2\pi {\tau _{LoS}}(t)f}}\\ 
&+ \sum\limits_{n = 1}^N {\sum\limits_{{l_n} = 1}^{{L_n}} {h_{pq,{l_n}}^{NLoS}(t){e^{ - j2\pi ({\tau _n}(t) + {\tau _{{l_n}}}(t))f}}} }.
\end{split}
\end{equation}
\subsection{Space-Time-Frequency Correlation Function}

In order to investigate the correlation of UHST channels, the space-time-frequency correlation function (STFCF) is calculated as 
\begin{equation}
\begin{split}
&{R_{pq,p'q'}}({\delta _p},{\delta _q},\Delta f,\Delta t;t,f)\\
&= E[{H_{pq}}(t,f)H_{p'q'}^*(t + \Delta t,f + \Delta f)]
\end{split}
\label{STFCF}
\end{equation}
where ${\delta _p} = \left\| {{\bf{A}}_p^T - {\bf{A}}_{p'}^T} \right\|$, ${\delta _q} = \left\| {{\bf{A}}_q^R - {\bf{A}}_{q'}^R} \right\|$. Due to the LoS component is determined based on Tx and Rx's relative position while NLoS components are determined based on parameters which are randomly generated, for simplicity, it is assumed here that the LoS and NLoS components are uncorrelated \cite{ref13}. Then (\ref{STFCF}) can be written as
\begin{equation}
\begin{split}
&{R_{pq,p'q'}}({\delta _p},{\delta _q},\Delta f,\Delta t;t,f)\\
&= R_{_{pq,p'q'}}^{LoS}({\delta _p},{\delta _q},\Delta f,\Delta t;t,f)
 + R_{_{pq,p'q'}}^{NLoS}({\delta _p},{\delta _q},\Delta f,\Delta t;t,f).
\end{split}
\end{equation}
The correlation function of the LoS component is expressed as follows [10]:
\begin{equation}
\begin{split}
&R_{pq,p'q'}^{LoS}({\delta _p},{\delta _q},\Delta f,\Delta t;t,f)\\
& = \frac{K}{{K + 1}}H_{pq}^{LoS}(t,f) \cdot H_{p'q'}^{LoS*}(t + \Delta t,f + \Delta f) .
\end{split}
\end{equation}
For NLoS components, the correlation function is expressed as
\begin{equation}
\begin{split}
&R_{pq,p'q'}^{NLoS}({\delta _p},{\delta _q},\Delta f,\Delta t;t,f) \\
&= \frac{1}{{K + 1}}\sum\limits_{n = 1}^N {\sum\limits_{{l_n} = 1}^{{L_n}} {H_{pq}^{NLoS}(t,f) \cdot H_{p'q'}^{NLoS*}(t + \Delta t,f + \Delta f)} } .
\end{split}
\end{equation}
In STFCF, let $\Delta f = 0 $, $q = q'$, $p = p' $, the function will be reduced to the time-variant ACF.
Let $\Delta t = 0 $, $\Delta f = 0 $, $q = q' $(or $p = p' $), the function will be reduced to the time-variant cross-correlation function (CCF) of the Rx (or Tx).
Let $\Delta t = 0 $, $q = q' $, $p = p' $, the function will be reduced to the time-variant frequency correlation function (FCF). 

\subsection{Stationary Interval}
The stationary interval is the minimum time interval during which the channel response remains constant. It can be used to determine the channel estimation frequency in ultra-high-speed mobile scenes \cite{ref10}. It is defined as the maximum length of time that the ACF of the power delay profile (PDP) exceeds a certain threshold ${\varsigma}$, namely, 
\begin{equation}
{\rm{I}} = \inf \{ \Delta t|{R_\Lambda }(t,\Delta t) \le \varsigma \}
\end{equation}

where ${\inf \{  \cdot \} }$ is the infimum of a function, ${{R_\Lambda }(t,\Delta t)}$ is the the ACF of the PDP and its calculation is given in \cite{ref13}. The threshold ${\varsigma}$ can be adjusted according to certain scenario and set to 80$\%$ here.

\section{Results and Discussions}
In this section, channel properties of the proposed channel model are simulated and analyzed. According to the size of existing vacuum tube train design of Hyperloop One \cite{ref20}, the cross section radius of vacuum tube is set $R$ = 2 m here. The material of tube wall is low carbon steel \cite{ref20} which can be approximated as a smooth surface, therefore the roughness is set ${\sigma _h} = 0 $ here. In metal tube, the position coordinates of the Tx is set as $({x_T},{y_T},{z_T}) = (0,0,4) $ while the initial position coordinates of the Rx is set as $({x_R},{y_R},{z_R}) = ({D_0},0,3) $, where $D_0 $ is the initial distance. At the Rx and Tx, a $2 \times 2$ MIMO linear antenna array communication systems are taken into consideration and antenna spacing are set as $\Delta {x_T} = \Delta {x_R} = \lambda  $ \cite{ref10}. Also, other parameters like carrier frequency ${f_c} = 58$ GHz and the train speed ${v^R} = 1080$ km/h. In the calculation of ${P_T}(\Delta t) $, the generation and recombination rate are set as ${r_b} = 80/{\rm{m}} $, ${r_d} = 4/{\rm{m}} $, respectively. The remaining parameters are randomly generated with reference to the 5G general channel model \cite{ref13}, and the equal area method (EAM) \cite{ref10} is used to obtain discrete AoA and AoD angle characteristics in channel model.
\subsection{The Time-Variant Spatial CCF}
\begin{figure}[tb]
	\centerline{\includegraphics[width=0.42\textwidth]{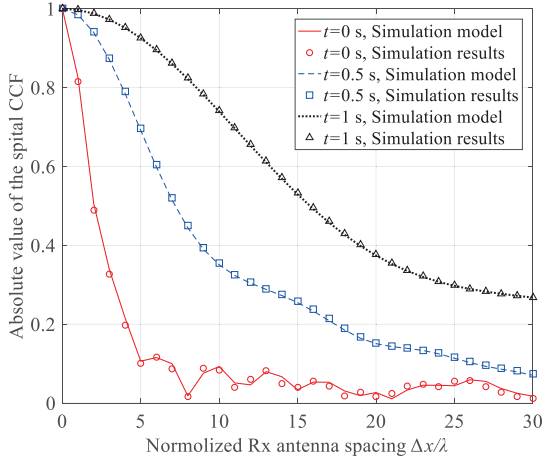}}
	\caption{Spatial CCFs comparison of simulation model and simulation results at different time instants ($R=2 $ m, ${\sigma _h} = 0 $, ${r_b} = 80/{\rm{m}} $, ${r_d} = 4/{\rm{m}} $, $D_0=600$ m, ${v^R} = 1080$ km/h, ${f_c} = 58 $ GHz, $k_1=k_2=6$).}
	\label{fig_CCF}
\end{figure}

The spatial CCFs comparison of the simulation model and simulation results at different time instants are shown in Fig.~\ref{fig_CCF}. Since the parameters are time-variant, such as azimuth AoA and elevation AoA, the spatial cross-correlation characteristics are different at different time instants, and also, the simulation model and simulation results curve fit well.

\subsection{The Time-Variant ACF}
\begin{figure}[tb]
	\centerline{\includegraphics[width=0.42\textwidth]{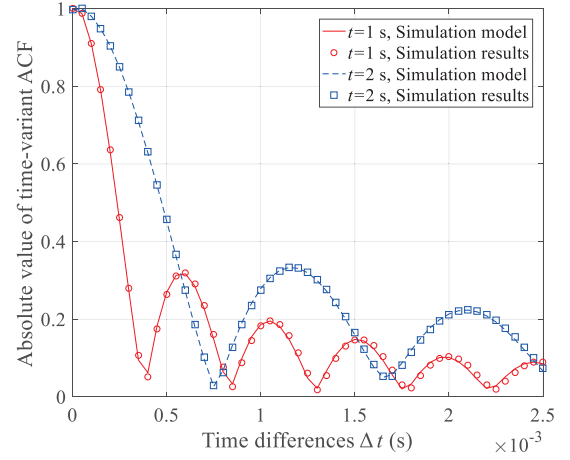}}
	\caption{ACFs comparison of simulation model and simulation results at different time instants ($R=2 $ m, ${\sigma _h} = 0 $, ${r_b} = 80/{\rm{m}} $, ${r_d} = 4/{\rm{m}} $, $D_0=600$ m, ${v^R} = 1080$ km/h, ${f_c} = 58$ GHz, $k_1=k_2=6$).}
	\label{fig_ACF}
\end{figure}

\begin{figure}[tb]
	\centerline{\includegraphics[width=0.42\textwidth]{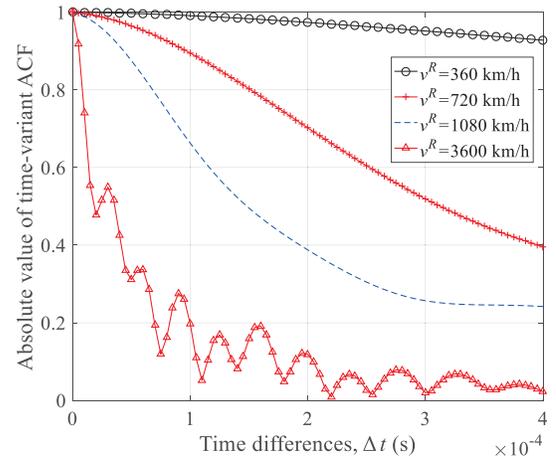}}
	\caption{Comparisons of ACFs of the simulation model for different ${v^R}$ at $t=0$ s ($R=2 $ m, ${\sigma _h} = 0 $, ${r_b} = 80/{\rm{m}} $, ${r_d} = 4/{\rm{m}} $, $D_0=600$ m, ${f_c} = 58$ GHz, $k_1=k_2=6$).}
	\label{fig_ACF_different_v}
\end{figure}

The ACFs comparison of simulation model and simulation results at different time instants are illustrated in Fig.~\ref{fig_ACF}. The curve fit of the simulation model and the simulation result is very good. The comparisons of ACFs of the simulation model for different ${v^R}$ at $ t=0$ s are shown in Fig.~\ref{fig_ACF_different_v}. As train speed increases, the ACF downward trend accelerates, and the attenuation is more rapid. In the future UHST scenarios, trains can reach thousands of kilometers per hour, which means the smaller coherence time will be considered in UHST channels.



\subsection{Comparison with Existing HST and Tunnel Channels}
In tunnel scenarios, the material of tunnel wall is generally reinforced concrete. Here, the roughness is set ${\sigma _h} = 0.002$ \cite{ref21} to simulate tunnel environment. The HST channel model in \cite{ref9} is used to modeling existing HST channel here. Some channel characteristics compared in above scenarios are shown as follows.

\begin{figure}[tb]
 	\centerline{\includegraphics[width=0.40\textwidth]{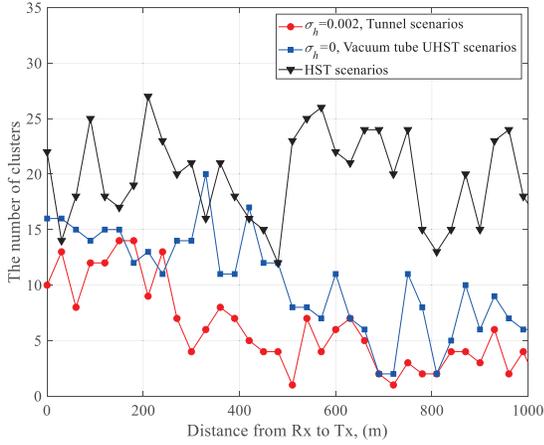}}
 	\caption{Comparisons of the number of clusters in different scenarios ($D_0=1000 $ m, ${v^R} = 1080 $ km/h, ${f_c} = 58 $ GHz, $k_1=k_2=6$).}
 	\label{fig_ClsterNum}
 \end{figure}

\begin{figure}[tb]
	\centerline{\includegraphics[width=0.42\textwidth]{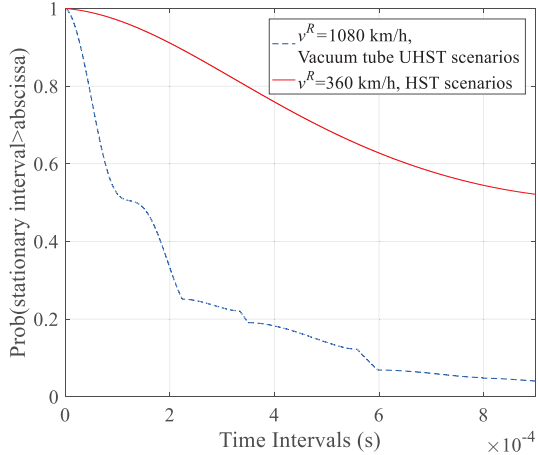}}
	\caption{Comparisons of empirical CCDFs of stationary intervals in different scenarios ($D_0=600 $ m, ${f_c} = 58 $ GHz, $k_1=k_2=6$).}
	\label{fig_Stationary_interval}
\end{figure}

The number of clusters changed with distance in three channels are illustrated in Fig.~\ref{fig_ClsterNum}. Due to extremely small space environment of the vacuum tube and tunnel, the number of clusters in their channels is much less than that in HST channels, and the same phenomenon in tunnel wireless communication can be found in \cite{ref12}. Moreover, compared with nearly smooth surface of vacuum tube, the signal will experience greater loss after passing through the rougher tunnel wall, which will reduce the effective propagation path to the Rx \cite{ref11}. Fig.~\ref{fig_Stationary_interval} compares the stationary intervals of vacuum tube UHST scenarios with existing HST scenarios. In existing HST scenarios, at a train speed of 360 km/h in 58 GHz, the stationary interval is about 0.35 ms, while it should be consider smaller in UHST channel, which is about 0.05 ms.

\section{Conclusions}
In this paper, a 3D non-stationary mmWave channel model for vacuum tube UHST communication systems has been proposed and its channel statistical properties have been studied, including time-variant ACF and spatial CCF. The results show that the non-stationarity of UHST channels. The simulation results match the simulation model well.  Moreover, by comparing channel properties of vacuum tube UHST scenarios with existing tunnel and HST scenarios, it is found that there are more multipaths in vacuum tube UHST channel than tunnel channels but less than HST channels. For future work, more statistical properties of UHST channel need to be investigated and the available measured date also need to be considered to modify the model once there are some channel measurements on vacuum tube UHST scenarios.


\section*{Acknowledgment}

\footnotesize {This work was supported by the National Key R\&D Program of China under Grant 2018YFB1801101, the China Railway Eryuan Engineering Group Co. Ltd Project under Grant KYY2019110(19-21), the National Natural Science Foundation of China (NSFC) under Grant 61960206006, the Frontiers Science Center for Mobile Information Communication and Security, the High Level Innovation and Entrepreneurial Research Team Program in Jiangsu, the High Level Innovation and Entrepreneurial Talent Introduction Program in Jiangsu, the Research Fund of National Mobile Communications Research Laboratory, Southeast University, under Grant 2020B01, the Fundamental Research Funds for the Central Universities under Grant 2242020R30001, the EU H2020 RISE TESTBED2 project under Grant 872172, and the Open Research Fund of National Mobile Communications Research Laboratory, Southeast University under Grant 2021D05.}


\end{document}